\documentclass[12pt]{iopart}
\usepackage{iopams} 

\def\nonu{\nonumber}
\def\br{\begin{eqnarray}}
\def\er{\end{eqnarray}}
\def\be{\begin{equation}}
\def\ee{\end{equation}}
\def\bse{\numparts}
\def\ese{\endnumparts}
\eqnobysec
\def\numparts{\refstepcounter{equation}%
     \setcounter{eqnval}{\value{equation}}%
     \setcounter{equation}{0}%
     \def\theequation{\arabic{section}.\arabic{eqnval}{\it\alph{equation}}}}
\def\endnumparts{\def\theequation{\arabic{section}.\arabic{equation}}%
     \setcounter{equation}{\value{eqnval}}}

\def\({\left(}
\def\){\right)}
\def\[{\left[}
\def\]{\right]}

\newcommand{\hepth}[1]{{\tt hep-th/#1}}

\def\bb{\mathbb}

\def\a{\alpha}

\def\d{\delta}
\def\D{\Delta}
\def\da{\dagger}

\def\what{\widehat}

\def\G{\Gamma}

\def\lm{\lambda}

\def\Om{\Omega}
\def\p{\phi}
\def\pt{{\widetilde \phi}}
\def\P{\Phi}
\def\pa{\partial}

\def\s{\sigma}
\def\S{\Sigma}
\def\t{\tau}

\def\tp0{\Theta_{+}^{(0)}}
\def\tm0{\Theta_{-}^{(0)}}

\def\vp{\varphi}
\def\vt{{\tilde \varphi}}

\def\psit{{\widetilde \psi}}
\def\Psit{{\widetilde\Psi}}

\def\f#1#2#3 {f^{#1#2}_{#3}}

\def\win1{{\sf w_{1+\infty}}}

\def\Win1{{\sf W_{1+\infty}}}

\newcommand{\poisson}[2]{\{#1\hspace{2.0pt}
 \mbox{\raisebox{2pt}{$\otimes$}}\hspace{-5.85pt}
 \mbox{\raisebox{-3pt}{,}}\hspace{5pt}#2\}}
\begin{document}

\title[Inverse scattering approach for Thirring models with defects]{Inverse scattering approach for massive Thirring models with integrable type-II defects}

\author{Alexis Roa Aguirre}

\address{Instituto de F\'\i sica Te\'orica - IFT/UNESP,\\
Rua Dr. Bento Teobaldo Ferraz, 271, Bloco II,
CEP 01140-070, S\~ao Paulo, SP, Brazil.}
\ead{aleroagu@ift.unesp.br}

\begin{abstract}
We discuss the integrability of the Bosonic and Grassmannian massive Thirring models in the presence of defects through the inverse scattering approach. We present a general method to compute the generating functions of modified conserved quantities for any integrable field equation associated to the $m \times m$ spectral linear problem. We apply the method to derive in particular the defect contributions for the number of occupation, energy and momentum of the massive Thirring models.
\end{abstract}


\section{Introduction} \label{sec:intro}

There has been a great deal of progress in studying two-dimensional integrable field theories in the presence of defects or internal boundaries, in both classical and quantum context \cite{Del,Sale1}.  From the classical point of view, the Lagrangian formalism \cite{Corr1,Bow2} has shown to be significantly successful. In this framework, the usual variational principle from a local Lagrangian density located at some fixed point, reveals frozen B\"acklund transformations as the defect conditions for the fields. There are many interesting features of these defects. It turns out that these kind of defect conditions allow for several types of integrable field theories \cite{Corr1,Bow2,Corr2}, not only the energy conservation but also the conservation of a modified momentum, which includes a defect contribution. Moreover, their integrability is provided by the existence of a modified Lax pair involving a limit procedure, but in general it was only checked explicitly for a few conserved charges. As a novel feature of most of these models is that only physical fields were present within the formulation and therefore the associated B\"acklund transformations were called type-I \cite{zambon1}. However, it was noticed that not all the possible relativistic integrable models could be accommodate within this framework and then it was proposed a generalization by allowing a defect to have its own degree of freedom, and the associated B\"acklund transformations were named type-II \cite{zambon1}. Many examples were also discussed in \cite{zambon1} like sine/sinh-Gordon, Liouville, massive free field and Tzitz\'eica-Bullogh-Dodd model. Concerning type-II defects, it is interesting to note that for the supersymmetric extensions of sine-Gordon model \cite{Lean1,Lean2} and for the massive Thirring models \cite{Ale,Ale2} those auxiliary boundary fields, which correspond to the degree of freedom of the defect itself, had already appeared naturally.
 
On the other hand, recently it was suggested an alternative and systematic new approach to defects in classical integrable field theories \cite{Cau}. The inverse scattering method formalism is used and the defect conditions corresponding to frozen B\"acklund transformations are encoded in a well-known defect matrix. This matrix provided an elegant way to compute the modified conserved quantities, ensuring integrability. Using this framework the generating function for the modified conserved charges for any integrable evolution equation of the AKNS scheme were computed, and the type-II B\"acklund transformations for the sine-Gordon and Tzitz\'eica-Bullough-Dodd models has been also recovered \cite{Ale3}. It is worth noting that this method for constructing integrable initial boundary value problems based on the B\"acklund transformations has already been used in \cite{Khabi,Tarasov}.

The aim of this paper is to provide an alternative approach in order to establish the integrability of the Bosonic and Grassmannian Thirring models with type-II defects, which were suggested by previous approaches \cite{Ale,Ale2}. At this point it is worth mentioning that there exists no totally clear definition of what integrability is for classical systems with infinitely many degrees of freedom. However, in this work we adopt the popular point of view where a system is regarded as integrable, if for its describing equations of motion it is possible to determine a constructive way of finding solutions and to show the existence of sufficient number of conserved quantities. Even though this viewpoint is not complete, it is sufficient for our immediate purposes. Since soliton solutions for these models have already been studied in \cite{Ale2}, herein we will focus on the explicit construction of their conserved quantities. 

The paper is organized as follows. In section 2, for the sake of clearness we present the standard setting of the Lax pair approach for the general $m\times m$ spectral linear problem (see for example \cite{AKNS}). One of the most important results of this approach and that is the main point for our purposes is the identification of sets of coupled Riccati equations in order to construct conservation laws by a simple generalization of the method used by Wadatti and Sogo for the particular case $m=2$ \cite{Wad1}. Other different generalizations, considering particularly matrix Riccati-type equations, have also been appeared in the literature concerning to multicomponent systems (see for example \cite{Tsu1} and references therein).

Within this framework, for the simplest case $m=2$ the defect contributions to the infinite set of modified integrals of motion were explicitly derived for several integrable equations associated with the ${\mathfrak{sl}(2)}$ Lie algebra-valued Lax pair like: (m)KdV, NLS, Liouville  and sine/sinh-Gordon equations for which previous results derived from Lagrangian principles \cite{Corr1, Corr2} were also recovered. More recently, the modified energy and momentum for the Tzitz\'eica model with type-II defects, which can be described by a $A_2^{(2)}$ Lie algebra-valued Lax pair, were also explicitly computed \cite{Ale3} and showed complete agreement with the results obtained by Lagrangian approach \cite{zambon1}. Then, motivated by these results, in section 3 we derive a general formula to compute the defect contributions to the infinite sets of modified conserved quantities for any $m\times m$ linear problem.

In section 4,  firstly we use this formula to compute explicitly the defect contributions to the modified energy and momentum for the Bosonic Thirring model. These significant results shed an interesting new light on the question of what type of B\"acklund transformations may be used as defect conditions since its Lagrangian description has not been derived until now. Subsequently, in section 5 we extend the procedure to include the pure fermionic version of the Thirring model which is described by ${\mathfrak sl}(2,1)$ algebra-valued Lax pair. We recursively compute the defect contributions to the modified energy and momentum and show that the results are in full agreement with the ones previously obtained by the Lagrangian formalism \cite{Ale}.

In section 6, we discuss in some detail how the question of involutivity of the charges can be addressed and how a Hamiltonian formulation can be provided to include type-II B\"acklund transformations which carry degrees of freedom corresponding to the defect itself.

Some concluding remarks are made in section 7. In Appendix A, we give a very short review of the sine-Gordon model with type-I defects within the framework adopted in this paper.

\section{Lax pair approach}

In this section we want to discuss the main ideas of the Lax formulation in order to construct an infinite set of independent conserved quantities for some integrable evolution equations. Such equations can be formulated as a compatibility condition of an associated linear auxiliary problem as follows,
\bse
\br 
  \pa_t \Psi(x,t;\lm) = V(x,t;\lm) \, \Psi(x,t;\lm),\label{s2e0.1}\\
 \pa_x \Psi(x,t;\lm) = U(x,t;\lm)  \,\Psi(x,t;\lm),\label{s2e0.2}
\er
\ese
where $\Psi(x,t;\lm)$ is an $m$-dimensional vector, $\lm$ is a spectral parameter, and $U,V$ are $m\times m$ matrices, which usually are named Lax pair. Then, from the compatibility condition, $\pa_x\pa_t \Psi(x,t;\lm)=\pa_t \pa_x\Psi(x,t;\lm)$, we obtain the zero-curvature equation,
\br
 \pa_t U - \pa_x V  + \left[U,V\] =0,\label{s2e0.3}
\er
which gives the corresponding equations of motion for the integrable system. Now, let us show how to construct a generating function for the infinite set of conservation laws. Firstly, for every auxiliary field component $\Psi_j$ with $j=1,...,m$, we can define a set of $(m-1)$ auxiliary functions $\G_{ij} = \Psi_{i}\Psi_j^{-1}$ with $i\neq j$. Then, considering the linear system (\ref{s2e0.1},) and (\ref{s2e0.2}), it is not so difficult to identify the $j$-th conservation equation,
\br
 \pa_t\left[U_{jj} + \sum_{i\neq j} U_{ji} \,\G_{ij}\] = \pa_x\left[V_{jj} + \sum_{i\neq j} V_{ji}\, \G_{ij}\],\label{s2e0.4}
\er
where each auxiliary functions $\G_{ij}$ satisfy coupled Riccati equations for the $x$-part,
\br
 \pa_x \G_{ij} = \(U_{ij}-U_{jj}\G_{ij}\) + \sum_{k\neq j}\bigg[U_{ik} - \G_{ij} \,U_{jk}\bigg] \G_{kj} \,,\label{equa2.4}
\er
and respectively for the $t$-part,
\br
  \pa_t \G_{ij} = \(V_{ij}-V_{jj}\G_{ij}\) + \sum_{k\neq j}\bigg[V_{ik} - \G_{ij} V_{jk}\bigg] \G_{kj},\label{s2e0.6}
\er
where without loss of generality we have assumed that $\Psi_j$ is a commuting field, however we will show later that this procedure also works in the case of anticommuting fields. 

Now, by considering solutions that vanish rapidly as $|x|\to \infty$, we found that the corresponding $j$-th generating function of the conserved quantities reads,
\br
 {\tt I}_j = \int_{-\infty}^{\infty}\rmd x\,\left[U_{jj} + \sum_{i\neq j} U_{ji} \,\G_{ij}\]. \label{s2e0.7}
\er

A wide group of integrable nonlinear evolution equations can be formulated using this approach, among which the most of known examples correspond to the particular case $m=2$, e.g, the nonlinear Schr\"odinger equation (NLS), Korteweg-de Vries (KdV) and the modified KdV equation (mKdV), Liouville equation, and sine/sinh-Gordon. For a more complete review of these cases see for example \cite{Segur}.

It is worth noting that if the respective analytic properties of the solutions are considered, we can expand the functions $\G_{ij}$ in positive and negative powers of $\lm$ and then solve (\ref{equa2.4}) and (\ref{s2e0.6}) recursively for each coefficient. This immediately provides an expansion of the $j$-th generating function $I_j$ in powers of $\lm$, obtaining in this way  an infinite set of conserved quantities. In particular, the usual conserved energy and momentum integrals of motion, commonly also derived from the Lagrangian formalism, turn out to be linear combinations of these set of conserved quantities $I_j$, by taking into account coefficients for the expansions in both positive and negative powers $\lm$. However, these sets of conserved quantities are not functionally independent in the bulk theory because not all of the auxiliary fields $\G_{ij}$ are independent.  Although, it seems that there is no need to consider all the conservation laws to derive the apparently overdetermined sets of conserved quantities,  the recent results on type-II integrable defects \cite{Ale3} suggest that for obtaining the most general form for the defect potentials, it is necessary to consider all the information coming from the Lax pair, i.e., all the conservation laws. To make it clearer, in the following section we will derive the formula for obtaining the modified conserved quantities which helps us to compute integrable defect potentials.


\section{Modified conserved charges from the defect matrix}

In this section we consider the Lax pair approach for constructing the infinite sets of modified conserved quantities in the presence of defects. 

Firstly, let us consider a defect placed at $x=0$, and suppose that there are two column-vector functions $\Psit$ and $\Psi$ corresponding to the auxiliary linear problems for $x<0$ described by the Lax pair $\widetilde{U}$, $\widetilde{V}$, and for $x>0$ by $U$ and $V$. Then, let us introduce $K(x,t;\lm)$, a matrix polynomial of the spectral parameter $\lm$, to connect the two solutions, namely,
\br
 \Psit(x,t;\lm) &=&  K(x,t;\lm)\,\Psi(x,t;\lm), \label{s3e0.1}
\er
where, $K$ satisfies differential equations corresponding to a gauge transformation \cite{Wad1} as follows,
\br
  \pa_t K = \widetilde{V}K - KV, \label{s3e0.2} \qquad 
  \pa_x K = \widetilde{U}K - UV,
\er
and it is commonly named the \emph{defect matrix} \cite{Corr1,Bow2,Corr2,Cau}. This matrix is expected to generate the defect conditions and consequently the corresponding auto-B\"acklund transformation of the model. A classification of these defect matrices was performed as well as several examples corresponding to the $m=2$ linear problem were examined by choosing a very simple form for this matrix \cite{Cau}. 

We present now a straightforward extension in order to consider the $m \times m$ matrix auxiliary linear problem.  Let us consider the generating functions (\ref{s2e0.7}) in the presence of the defect,
\br
 {\cal I}_j = \int_{-\infty}^0 \rmd x\,\left[\widetilde{U}_{jj} + \sum_{k\neq j} \widetilde{U}_{jk} \widetilde{\G}_{kj} \] + \int_0^{\infty}\rmd x\,\left[U_{jj} + \sum_{k\neq j} U_{jk} \G_{kj} \].
\er
Hence, taking the time derivative and using the conservation equation (\ref{s2e0.4}), we get
\br
 \frac{\rmd {\cal I}_j}{\rmd t} &=&\left[\widetilde{V}_{jj} + \sum_{i\neq j} \widetilde{V}_{ji} \widetilde{\G}_{ij}\]\Bigg|_{x=0} -\left[V_{jj} + \sum_{i\neq j} V_{ji} \G_{ij}\]\Bigg|_{x=0}. \label{s3e0.5}
\er
Then, it is not difficult to show that the relation between the two sets of auxiliary functions $\widetilde{\G}_{ij}$ and $\G_{ij}$ is given by
\br
  \widetilde{\G}_{ij} &=& \left[\frac{K_{ij}+\sum\limits_{k\neq j}K_{ik}\,\G_{kj}}{K_{jj} + \sum\limits_{k\neq j} K_{jk}\,\G_{kj}}\right].
\er
Inserting in (\ref{s3e0.5}), one gets
\br
 \frac{\rmd{\cal I}_j}{\rmd t} &=& \frac{\Om_j}{\D_j}, \qquad \mbox{where} \qquad  \D_j\,=\,K_{jj} + \sum\limits_{k\neq j} K_{jk}\,\G_{kj},
\er
and
\br
 \fl \Om_j \!\!&=& \!\!\!\(\widetilde{V}_{jj} - V_{jj} -\sum_{i\neq j} V_{ji} \,\G_{ij} \)\D_j  + \sum_{i\neq j} \widetilde{V}_{ji} K_{ij}  + \sum_{i,k \neq j} \widetilde{V}_{ji} K_{ik} \,\G_{kj}. \qquad \mbox{}
\er
Finally, we consider the equations (\ref{s2e0.6}) and (\ref{s3e0.2}) to obtain,
\br
 \frac{\rmd}{\rmd t} \left\{{\cal I}_j-\ln\left[ K_{jj} + \sum\limits_{k\neq j} K_{jk}\G_{kj}\right]\Bigg|_{x=0}\right\} =0 , \label{s3e0.39}
\er
where the defect contribution to the $j$-th generating function of infinite conserved quantities is given exactly by 
\br 
D_j = -\ln\left[ K_{jj} + \sum\limits_{k\neq j} K_{jk}\G_{kj}\right]\Bigg|_{x=0} . \label{formu}
\er 
This formula is an important result because its expansion in powers of $\lm$ provides the defect contributions to the modified conserved quantities at all orders for every $m \times m$ associated linear problem.  
We will apply it to study the Bosonic Thirring model, which is described by a ${\mathfrak sl}(2)$ Lie algebra-valued Lax pair, and for the Grassmannian Thirring model, which is associated with a ${\mathfrak sl}(2,1)$ Lie algebra-valued Lax pair. In particular, it will be shown that the modified energy and momentum contributions can be computed from certain linear combinations of the set of conserved quantities ${D_j}$, taking into account all the possible conservation laws, i.e., for $j=1,...,m$. It is worth noting that a similar approach was already used in \cite{Hab2} to prove the classical and quantum integrability of the sine-Gordon model with defects, by using the monodromy matrix language. In the same work all higher conserved quantities are found by using a matrix B\"acklund tranformation and a matrix Riccati equation. Some of these results are recovered using our approach in {Appendix A.}


\section{Case $m=2$ : The Bosonic Thirring model}

In this section, we apply the method described by computing explicitly the defect contributions to the modified energy and  momentum for the BT model.

\subsection{The bulk BT model and the linear problem}

As is well known the BT model in the bulk is integrable \cite{Kuz} and the associated linear problem can be formulated by using $2\times 2$ matrices valued in the ${\mathfrak sl}(2)$ algebra as follows,
\bse
\br 
  \pa_t \Psi(x,t;\lm) = V(x,t;\lm) \, \Psi(x,t;\lm),\label{s2e1}\\
 \pa_x \Psi(x,t;\lm) = U(x,t;\lm)  \,\Psi(x,t;\lm),\label{s2e2}
\er
\ese
where the auxiliary field $\Psi =(\Psi_1, \Psi_2)^T$ is a 2-vector and the Lax pair can be written in a compact form as,
\bse
\br
  U &=&
 \left[
   \begin{array}{cc} \frac{\rmi}{4}\left[g\rho_- -m\big(\lm^2 -\lm^{-2}	\big)\]   & q(\lm) \\[0.3cm]
    r(\lm) &   -\frac{\rmi}{4}\left[g\rho_- -m\big(\lm^2 -\lm^{-2}	\big)\] 
   \end{array}
 \], \\[0.2cm]
 V &=&
 \left[
   \begin{array}{cc} -\frac{\rmi}{4}\left[g\rho_+ +m\big(\lm^2 +\lm^{-2}	\big)\]  & B(\lm) \\[0.3cm]
   C(\lm) & \frac{\rmi}{4}\left[g\rho_+ +m\big(\lm^2 +\lm^{-2}	\big)\] 
   \end{array}
 \],\qquad \mbox{}
\er
\ese
where for convenience we have defined $\rho_{\pm} \,=\,(\p_2^{\da}\p_2 \pm \p_1^\da\p_1 )$, and the following functions,
\bse
\br
  B(\lm) &= \frac{\rmi\sqrt{mg}}{2} \big(\lm\p_1 - \lm^{-1} \p_2 \big), \quad  & q(\lm) =\, \frac{\rmi\sqrt{mg}}{2} \big(\lm\p_1 + \lm^{-1} \p_2 \big),  \\[0.1cm]
  C(\lm) &= -\frac{\rmi\sqrt{mg}}{2} \big(\lm \p_1^\da - \lm^{-1} \p_2^{\da} \big), \quad &r(\lm) =\, -\frac{\rmi\sqrt{mg}}{2} \big(\lm \p_1^\da +\lm^{-1} \p_2^{\da} \big).
\er
\ese
From the zero curvature condition, we obtain the field equations for the BT model 
\bse
\br
 \rmi(\pa_t -\pa_x) \p_1 &=& m \p_2 + g\p_2^{\da}\p_2\p_1,\\
\rmi(\pa_t + \pa_x) \p_2 &=& m\p_1 + g\p_1^{\da}\p_1\p_2,\\
 \rmi(\pa_t -\pa_x) \p_1^{\da} &=& -m\p_2^{\da} -g\p_1^{\da}\p_2^{\da}\p_2,\\
\rmi(\pa_t + \pa_x) \p_2^{\da} &=& -m\p_1^{\da}-g\p_2^{\da}\p_1^{\da}\p_1.
 \er
\ese
Now, we define the auxiliary function $\G_{21}= \Psi_2\Psi_1^{-1}$. Then, by using the system of linear equations we have that the conservation equation can be written in the following form,
\br
 \pa_t\[q\G_{21} + \frac{\rmi g}{4}\rho_- \] = \pa_x\bigg[B\G_{21} -\frac{\rmi g}{4}\rho_+  \bigg].
\er
The auxiliary function $\G_{21}$ satisfies the following Riccati equations
\bse
\br
 \pa_x \G_{21} &=& r -\frac{\rmi}{2}\left[g\rho_- -m\(\lm^2 -\lm^{-2}\) \] \G_{21} - q \G_{21}^2 ,\\[0.1cm]
 \pa_t \G_{21} &=& C + \frac{\rmi}{2}\left[g\rho_+ +m\(\lm^2 +\lm^{-2}\) \] \G_{21} -B\G_{21}^2.
\er
\ese
Now, we expand $\G_{21}$ in inverse powers of $\lm$ around $\infty$,
\br
 \G_{21}(x,t;\lm) = \sum_{k=0}^{\infty} \frac{\G_{21}^{(k)}(x,t)}{\lm^{k}}\,.\label{s2.2.9}
\er
Using the Riccati equation, each expansion coefficient ${\G_{21}^{(k)}(x,t)}$ can be obtained easily in a recursive way. The first coefficients are given by
\br
\fl
 {\G_{21}^{(1)}}= \sqrt{\frac{g}{m}}\,\p_1^\da, \quad {\G_{21}^{(2)}}\,=\,0, \quad {\G_{21}^{(3)}}\,=\, \sqrt{\frac{g}{m}}\left[-\frac{2\rmi}{m} (\pa_x\p_1^\da)  + \p_2^\da +\frac{g}{m}(\p_2^\da\p_2)\p_1^\da \].
\er
Considering, as usual, the bosonic fields $\p_i(x,t)$ vanish at $|x|\to \infty$, the corresponding generating function for the conserved quantities reads
\br
 {\tt I}_1 &=& \int_{-\infty}^{\infty}\rmd x\,\left[q\G_{21} + \frac{\rmi g}{4}\rho_-\],\label{s2.e2.11}
\er
and substituting (\ref{s2.2.9}) in the expression for ${\tt I}_1$, we get an infinite number of conserved quantities given by the expansion
\br
 {\tt I}_1 = \sum_{k=0}^{\infty} \frac{{\tt I}_1^{(k)}}{\lm^{2k}}.
\er
Then, the first two conserved quantities are explicitly given by
\bse\br 
 \fl {\tt I}_1^{(0)} = \frac{\rmi g}{4}\int_{-\infty}^{\infty} \rmd x \left[\p_1^\da\p_1 + \p_2^{\da}\p_2 \],\label{s4e14}\\[0.3cm]
\fl  {\tt I}_1^{(2)} = -\frac{\rmi g}{m} \int_{-\infty}^{\infty}\rmd x \left[\rmi\p_1(\pa_x\p_1^\da) - \frac{m}{2}\(\p_2^\da\p_1 + \p_1^{\da}\p_2 \) -\frac{g}{2}(\p_1^\da\p_1 \p_2^{\da}\p_2 ) \].\label{s4e15}
\er\ese
\noindent In addition, there is another set of conserved quantities that can be computed taking an expansion of $\G_{21}(x,t;\lm)$ in positive powers of $\lm$,
\br
 \G_{21}(x,t;\lm) = \sum_{k=0}^{\infty}{\what{\G}^{(k)}}_{21}(x,t)\,\lm^{k}.
\er
In a very similar way, the first coefficients are,
\br
\fl
  {\what \G}_{21}^{(1)} &=& -\sqrt{\frac{g}{m}}\,\p_2^\da, \quad {\what \G}_{21}^{(2)} \,=\,0, \quad {\what \G}_{21}^{(3)}\,=\, \sqrt{\frac{g}{m}}\left[-\frac{2\rmi}{m} (\pa_x\p_2^\da)  - \p_1^\da -\frac{g}{m}(\p_1^\da\p_1)\p_2^\da \].\qquad \mbox{}
\er
Substituting in (\ref{s2.e2.11}), we will now obtain that the conserved quantities read
\br
 {\tt I}_1 &=&  \sum_{k=0}^{\infty} {\tt{\what I}}_1^{(k)}\,\lm^{2k}\,,
\er
where, the first two of them have been computed  schematically, and the result is the following	
\bse\br
 \fl  {\what {\tt I}}_1^{(0)} &=&\!\! -\frac{\rmi g}{4}\int_{-\infty}^{\infty}\rmd x \left[\p_1^\da\p_1 + \p_2^{\da}\p_2 \],\label{s4e19}\\[0.3cm]
\fl {\what {\tt I}}_1^{(2)} &=&\!\! -\frac{\rmi g}{m} \int_{-\infty}^{\infty}\rmd x \left[\rmi\p_2(\pa_x\p_2^\da) + \frac{m}{2}\( \p_2^{\da}\p_1 +\p_1^\da\p_2\) +\frac{g}{2}(\p_1^\da\p_1 \p_2^{\da}\p_2 ) \].\label{s4e20}
\er\ese
Then, we have found two infinite set of independent conserved quantities as consequence of the two possible choices for the $\lm$-expansion of the auxiliary function $\G_{21}(x,t;\lm)$, i.e, around $\lm=0$ and $\lm=\infty$. 
However, these integrals of motion are not really real charges. Then, it is necessary to add their corresponding complex conjugate terms. In fact, these terms naturally raise by considering a second conservation equation that can be derived from the linear system (\ref{s2e1}) and (\ref{s2e2}), namely 
\br
 \pa_t\left[r\G_{12} -\frac{\rmi g}{4}\rho_- \] = \pa_x \left[C\G_{12} +\frac{\rmi g}{4}\rho_+ \],\label{s4e21}
\er
where we have introduced a new auxiliary function $\G_{12} =\Psi_1\Psi_2^{-1}$, which also satisfy a couple of Riccati equations,
\bse\br
 \pa_x \G_{12} &=& q +\frac{\rmi}{2}\left[g\rho_- -m\(\lm^2-\lm^{-2}\)\]\G_{12} -r\G_{12}^{2}\,,\\[0.1cm]
 \pa_t \G_{12} &=& B -\frac{\rmi}{2}\left[g\rho_+ +m\(\lm^2+\lm^{-2}\)\]\G_{12} -C\G_{12}^{2}.
\er\ese
Then, using the same scheme we can obtain recursively the first few coefficients for the auxiliary function $\G_{12}(x,t;\lm)$ by considering the corresponding expansion in negative and positive powers of $\lm$. Doing that, the results obtained are:
\bse\br
\fl
  {\G_{12}^{(1)}}&=& \sqrt{\frac{g}{m}}\,\p_1, \quad \,\,\,\,\,\,\, {\G_{12}^{(2)}}\,=\,0, \quad {\G_{12}^{(3)}}\,=\, \sqrt{\frac{g}{m}}\left[\frac{2\rmi}{m} (\pa_x\p_1)  + \p_2 +\frac{g}{m}(\p_2^\da\p_2)\p_1\],\\[0.1cm]
 \fl {\what \G}_{12}^{(1)} &=& -\sqrt{\frac{g}{m}}\, \p_2, \quad {\what \G}_{12}^{(2)} \,=\,0, \quad {\what \G}_{12}^{(3)}\,=\, \sqrt{\frac{g}{m}}\left[\frac{2\rmi}{m} (\pa_x\p_2)  - \p_1 -\frac{g}{m}(\p_1^\da\p_1)\p_2\].
\er\ese
From the conservation equation (\ref{s4e21}), the second generating function of the conserved quantities can be written as follows,
\br
 {\tt I}_2 = \int_{-\infty}^{\infty}\rmd x \left[r\G_{12} -\frac{\rmi g}{4}\rho_- \].\label{s4e26.}
\er
Substituting the corresponding coefficients of the auxiliary functions for each expansion in $\lm$, we obtained the following conserved quantities,
\bse
\br
 {\tt I}_2^{(0)} = -\frac{\rmi g}{4}\int_{-\infty}^{\infty}\rmd x \left[\p_1^\da\p_1 + \p_2^{\da}\p_2 \]\,\,=\,\, {\what{\tt I}}_2^{(0)} ,\\[0.3cm]
\fl {\tt I}_2^{(2)} = -\frac{\rmi g}{m} \int_{-\infty}^{\infty}\rmd x \left[\rmi\p_1^\da(\pa_x\p_1) + \frac{m}   {2}\(\p_2^\da\p_1 + \p_1^{\da}\p_2 \) +\frac{g}{2}(\p_1^\da\p_1 \p_2^{\da}\p_2 ) \], \\[0.3cm]
\fl {\what {\tt I}}_2^{(2)} = -\frac{\rmi g}{m} \int_{-\infty}^{\infty}\rmd x \left[\rmi\p_2^\da(\pa_x\p_2) - \frac{m}{2}\( \p_2^{\da}\p_1 +\p_1^\da\p_2\) -\frac{g}{2}(\p_1^\da\p_1 \p_2^{\da}\p_2 ) \].
\er
\ese
We note that the usual number of occupation, energy and momentum for the BT model can be expressed in the following form,{\small
\bse\br
\fl  N &=&\frac{1}{\rmi g}\left[({\tt I}_1^{(0)} - {\tt I}_2^{(0)}) - ({\what{\tt I}}_1^{(0)} -{\what{\tt I}}_2^{(0)} )\] \,=\, \int_{-\infty}^{\infty} \rmd x\,\left[\p_1^\da\p_1 + \p_2^{\da}\p_2 \]\,,\\
 \fl E &=&\frac{\rmi m}{2g} \left[({\tt I}_1^{(2)} - {\tt I}_2^{(2)}) - ({\what{\tt I}}_1^{(2)} - {\what{\tt I}}_2^{(2)} )\]  \\
 \fl &=&\int_{-\infty}^{\infty} \rmd x \left[\frac{\rmi}{2}\(\p_1\pa_x \p_1^{\da}-\p_1^{\da}\pa_x \p_1
       -\p_2\pa_x \p_2^{\da}+\p_2^{\da}\pa_x \p_2\) - m\big(\p_2^{\da}\p_1+
       \p_1^{\da}\p_2\big) -g\,\p_1^{\da}\p_1\p_2^{\da}\p_2  \right],\nonumber\\
\fl P&=&\frac{\rmi m}{2g} \left[({\tt I}_1^{(2)} - {\tt I}_2^{(2)}) + ({\what{\tt I}}_1^{(2)} -{\what{\tt I}}_2^{(2)} )\] \nonumber \\
\fl &=&\int_{-\infty}^{\infty}\rmd x\left[\frac{\rmi}{2}\( \p_1\pa_x \p_1^{\da} -\p_1^{\da} \pa_x \p_1
       + \p_2 \pa_x \p_2^{\da}-\p_2^{\da}\pa_x \p_2 \)\right].
\er\ese}\normalsize
In the next section we derive the defect contribution for the energy and momentum by using directly the defect matrix.

\subsection{Modified integrals of motion of the BT model from the defect matrix}

As it was shown in section 3, in order to compute the defect contribution to each bulk integral of motion, it is necessary to know the explicit form of the elements of the defect matrix. Using an ansatz for the expansion of this matrix as a very simple Laurent series,
\br
 K &=& K_{-1} +  K_0 + K_1 ,
\er
where $K_i$ corresponds to an element of grade $\lm^{i}$, a totally consistent defect matrix was determined \cite{Ale2}, which can be written in the following form
\br
  K &=& \left[
   \begin{array}{cc} 
   -\sqrt{\frac{m}{g}}\Big[ \lm \rme^{-\rmi\a} - \rmi (\lm a)^{-1} \rme^{\rmi\a} \Big] & X \\
   - X^{\da} & \sqrt{\frac{m}{g}}\left[\lm \rme^{\rmi\a} + \rmi (\lm a)^{-1}\rme^{-\rmi\a} \]
   \end{array}
 \right],\qquad \mbox{} \label{s4e26}
\er
where 
\br
 2 \a &=& \arcsin\[\frac{ga}{2m}X^\da X\],
\er
which it turns to be related to the modified number of occupation by $4\a = g N_{\rm D}$.
\noindent Here, the boundary fields $X$ and $ X^\da$ satisfy the following algebraic relations,
\bse
\br
 X &=& \pt_1  \rme^{\rmi\a} + \p_1  \rme^{-\rmi\a}  \,=\, \frac{\rmi}{a}\left[\p_2 \rme^{\rmi\a}  - \pt_2 \rme^{-\rmi\a} \],\label{s4e28} \\[0.1cm]
 X^{\dagger} &=&\pt_1^{\dagger} \rme^{-\rmi\a}  + \p_1^{\dagger}  \rme^{\rmi\a} \,=\, \frac{1}{\rmi a}\left[\p_2^{\dagger}  \rme^{-\rmi\a} - \pt_2^{\dagger}  \rme^{\rmi\a}  \] ,\label{ss4e29}
\er
\ese
and the respective time-derivatives,
\bse
\br
\pa_t X &=& \frac{m}{2a}\(\p_1 \rme^{\rmi\a}   - \pt_1 \rme^{-\rmi\a}   \) -\frac{\rmi m}{2}\(\pt_2  \rme^{\rmi\a}  + \p_2 \rme^{-\rmi\a} \)\nonumber \\
  &\mbox{}& -\frac{\rmi g}{4}\left[\pt_1^{\da}\pt_1 +\p_1^{\da}\p_1+\pt_2^{\da}\pt_2 +\p_2^{\da}\p_2\] X , \\[0.3cm]
  \pa_t X^\da &=& \frac{m}{2a}\(\p_1^{\da}  \rme^{-\rmi\a}  - \pt_1^{\da}  \rme^{\rmi\a}    \) +\frac{im}{2}\(\pt_2^{\da}\rme^{-\rmi\a}  + \p_2^{\da} \rme^{\rmi\a}  \)\nonumber \\
  &\mbox{}& +\frac{\rmi g}{4}\left[\pt_1^{\da}\pt_1 +\p_1^{\da}\p_1+\pt_2^{\da}\pt_2 +\p_2^{\da}\p_2 \] X^\da,
\er
\ese
together with,
\bse
\br
 \pa_x X &=& \frac{m}{2a}\(\p_1  \rme^{\rmi\a}  - \pt_1 \rme^{-\rmi\a}   \) +\frac{im}{2}\(\pt_2  \rme^{\rmi\a}  + \p_2  \rme^{-\rmi\a}  \)\nonumber \\
&\mbox{}& -\frac{\rmi g}{4}\left[\pt_1^\da\pt_1 +\p_1^{\da}\p_1-\pt_2^\da\pt_2 -\p_2^{\da}\p_2 \] X , \\[0.3cm]
 \pa_x X^\da &=& \frac{m}{2a}\(\p_1^{\da}  \rme^{-\rmi\a} - \pt_1^{\da} \rme^{\rmi\a}   \) -\frac{im}{2}\(\pt_2^{\da}  \rme^{-\rmi\a} + \p_2^{\da}  \rme^{\rmi\a}  \)\nonumber \\
  &\mbox{}& +\frac{\rmi g}{4}\left[\pt_1^{\da}\pt_1 +\p_1^{\da}\p_1-\pt_2^{\da}\pt_2 -\p_2^{\da}\p_2\] X^\da,\label{s4e33}
\er
\ese
where $a$ is a real parameter. The expressions (\ref{s4e28})--(\ref{s4e33}) are the auto-B\"acklund transformations for the BT model. It was also shown in \cite{Ale2} that these transformations exhibit a complete consistency with the soliton solutions derived by applying the Dressing method for several transitions between them. 

Once a defect matrix is given by (\ref{s4e26}), the defect contribution to the modified conserved quantities can be calculated using (\ref{formu}). Firstly, let us consider the generating function of conserved quantities (\ref{s2.e2.11}) in the presence of a defect,
\br
 {\cal I}_1 &=& \int_{-\infty}^{0}\rmd x\,\left[\widetilde{q} \,\widetilde{\G}_{21} + \frac{\rmi g}{4}\widetilde\rho_-\] +\int_{0}^{\infty}\rmd x\,\left[q\G_{21} + \frac{\rmi g}{4}\rho_-\].
\er
Then, taking the time-derivative and using the field equations we have,
\br
 \fl \frac{\rmd{\cal I}_1}{\rmd t} &=& \left[\widetilde{B} \widetilde{\G}_{21} -\frac{\rmi g}{4}\widetilde\rho_+\]\bigg|_{x=0} -\left[B\G_{21} -\frac{\rmi g}{4}\rho_+ \]\bigg|_{x=0}\quad \equiv \,\,\,\,\,- \frac{\rmd { D}_1}{\rmd t}\,, 
\er
with
\br
 {D}_1 &=& -\ln\left[ K_{11} +  K_{12}\G_{21}\right]\bigg|_{x=0}\,.
\er
Now, the methodology used before to obtain the conserved quantities in the bulk, can be applied directly. First, we calculate the conserved charges associated to the expansion of $\G_{21}$ in inverse powers of $\lm$. Then, we consider the corresponding expansion in positive powers of $\lm$. Implementing this procedure, we find out the following results,
\bse\br
D_1^{(2)} &=& \frac{\rmi}{a}\,\rme^{2\rmi\a}+ \frac{g}{m}X\p_1^{\da}\,\rme^{\rmi\a}, \\
 \widehat{D}_1^{(2)} &=& -\rmi a \, \rme^{-2\rmi\a} -\frac{\rmi ag}{m}X\p_2^{\da}\,\rme^{-\rmi\a}.
\er\ese
Performing the same procedure for the generating function (\ref{s4e26.}), one gets
\br
\fl  \frac{\rmd{\cal I}_2}{\rmd t} &=& \left[\widetilde{C}\widetilde{\G}_{12} +\frac{\rmi g}{4}\widetilde\rho_+ \]\bigg|_{x=0} -\left[C\,\G_{12} +\frac{ig}{4}\rho_+ \]\bigg|_{x=0}\quad \equiv \,\,\,\,\,- \frac{\rmd {D}_{2}}{\rmd t}\,, 
\er
where
\br
 D_2 &=& -\ln\left[ K_{22} +  K_{21}\G_{12}\right]\bigg|_{x=0}\,,
\er
and from which, we obtain the following coefficients,
\bse\br
 D_2^{(2)} &=& -\frac{\rmi}{a}\,\rme^{-2\rmi\a} + \frac{g}{m}X^\da\p_1\,\rme^{-\rmi\a}, \\
 \widehat{D}_2^{(2)}  &=&\rmi a \, \rme^{2\rmi\a} +\frac{\rmi ag}{m}X^\da\p_2\,\rme^{\rmi\a}.
\er\ese
By analogy we find that defect energy and momentum for the BT model can be written in the following way,\bse
\br
\fl 
 E_D &=& \frac{\rmi m}{2g} \left[(D_1^{(2)} -D_2^{(2)}) - (\widehat{D}_1^{(2)}-\widehat{D}_2^{(2)})\] \,=\, -\frac{m}{2g}\(a+\frac{1}{a}\) \(\rme^{2\rmi\a} +\rme^{-2\rmi\a}\) \nonu \\
\fl  &\mbox{}& +\frac{\rmi}{2}\(X\p_1^{\da}\,\rme^{\rmi\a} - X^\da\p_1\,\rme^{-\rmi\a}\) -\frac{a}{2}\(X\p_2^{\da}\,\rme^{-\rmi\a}+X^\da\p_2\,\rme^{\rmi\a}\), \qquad \mbox{}\\
\fl P_D &=& \frac{\rmi m}{2g} \left[(D_1^{(2)} -D_2^{(2)}) + (\widehat{D}_1^{(2)}-\widehat{D}_2^{(2)})\] \,=\, \frac{m}{2g}\(a-\frac{1}{a}\) \(\rme^{2\rmi\a} +\rme^{-2\rmi\a}\) \nonu \\
 \fl &\mbox{}& +\frac{\rmi}{2}\(X\p_1^{\da}\,\rme^{\rmi\a} - X^\da\p_1\,\rme^{-\rmi\a}\) +\frac{a}{2}\(X\p_2^{\da}\,\rme^{-\rmi\a} +X^\da\p_2\,\rme^{\rmi\a}\).
\er\ese
We can note that it is possible to rewrite these results in an alternative form by using the B\"acklund transformations  (\ref{s4e28})--(\ref{s4e33}), as follows
\bse\br
\fl  E_D &=&\frac{\rmi} {2}\left[\(\p_1^{\da}\pt_1-\pt_2^{\da}\p_2 \)\rme^{2\rmi\a} 			    	    			-\(\pt_1^{\da}\p_1-\p_2^{\da}\pt_2\)\rme^{-2\rmi\a}\] -\frac{m}{g}\(a+\frac{1}{a}\) \cos\(2\a\) , \\
\fl P_D &=&  \frac{\rmi} {2}\left[\(\p_1^{\da}\pt_1+\pt_2^{\da}\p_2 \)\rme^{2\rmi\a}			    	    			-\(\pt_1^{\da}\p_1+\p_2^{\da}\pt_2\)\rme^{-2\rmi\a}\] +\frac{m}{g}\( a- \frac{1}{a} \) \cos\(2\a\).
\er\ese
These expressions for the defect energy and momentum seem not to have been reported elsewhere in the literature and constitute a very important result in order to address in future works the question of the Lagrangian formalism as well as quantum aspects like the transmission matrix.

Since the integrable defect conditions  for the BT model have already been determined by giving the corresponding auto-B\"acklund transformations, the integrability of the model in the presence of defects, following the integrability criteria adopted in this work, is provided by the existence of the defect matrix and the explicit computations of the modified conserved quantities.


\section{Case $m=3$ : The Grassmanian Thirring model}

In this section, we will apply the formalism on the GT model. As it was done in the Bosonic case, we will construct an infinite set of independent conserved quantities for the bulk theory, and then derive the corresponding defect contributions to the modified conserved quantities.

\subsection{The bulk GT model and the linear problem}

The equations of motion for the GT model can be described as a compatibility condition of the following associated linear problem,
\bse
\br
 \pa_x \Psi(x,t;\lm) &=& U(x,t;\lm)\Psi(x,t;\lm),\label{e5.1}\\
 \pa_t \Psi(x,t;\lm) &=& V(x,t;\lm)\Psi(x,t ;\lm). \label{e5.2}
\er
\ese
where the three-wave function $\Psi$ has the form $(\Psi_1,\Psi_2,\Psi_3)^T$. We remark that the Bosonic or Grassmaniann property of these functions depends on the conservation law that will be considered. The Lax pair $U,V$ are $3\times 3$ matrices belong to the ${\mathfrak sl}(2,1)$ Lie algebra (see for example Appendix C in \cite{Ale3}), and can be explicitly written as follows,
\bse
\br
 \fl U &=&  \left[\begin{array}{ccc} 
                         \frac{\rmi g}{2}\,\rho_- +\frac{\rmi m}{2}\(\lm^2-\lm^{-2}\) & 0 & q_1(\lm) \\[0.1cm] 0 & -\frac{ig}{2}\,\rho_- +\frac{\rmi m}					                         {2}\(\lm^2-\lm^{-2}\)  & q_2(\lm) \\[0.1cm] r_1(\lm)  & r_2(\lm) & \rmi m(\lm^2 - \lm^{-2})
                      \end{array}
              \right], \\[0.3cm]
\fl V &=&  \left[\begin{array}{ccc} 
                        -A +\frac{\rmi m}{2}\(\lm^2+\lm^{-2}\) & 0 & B_1(\lm) \\[0.1cm] 0 & A +\frac{\rmi m}					                                	 				                 {2}\(\lm^2+\lm^{-2}\)  & B_2(\lm) \\[0.1cm] C_1(\lm)  & C_2(\lm) & \rmi m(\lm^2 + \lm^{-2})
                    \end{array}
              \right],            
\er
\ese
where just for simplicity, we have also defined  the functions
\br
\fl A &=& \frac{\rmi g}{2}\rho_+, \qquad \rho_{\pm} = (\psi_2^\da\psi_2 \pm\psi_1^\da\psi_1),\\
\fl  q_1 &=& -C_2\,=\,-\rmi\sqrt{\frac{mg}{2}} \(\lm\psi_1 +\lm^{-1}\psi_2\), \qquad q_2\,=\,-C_1\,=\, \rmi\sqrt{\frac{mg}{2}}\(\lm \psi_1^\da -\lm^{-1}\psi_2^\da \), \qquad \mbox{}\\
\fl  r_1 &=& -B_2 \,=\,-\rmi\sqrt{\frac{mg}{2}}\(\lm \psi_1^\da +\lm^{-1}\psi_2^\da \), \qquad\!
 r_2 \,=\, -B_1\,=\, \rmi\sqrt{\frac{mg}{2}}\(\lm \psi_1 -\lm^{-1}\psi_2\).
\er
By applying the zero-curvature condition we obtain the field equations for GT model,
\begin{eqnarray}
 \rmi(\pa_t -\pa_x) \psi_1 &=& m \psi_2+ g\psi_2^{\dagger}\psi_2\psi_1\,,\label{e1}\\
 \rmi(\pa_t + \pa_x) \psi_2 &=& m\psi_1 + g\psi_1^\da\psi_1\psi_2\,,\label{e2}\\
 \rmi(\pa_t -\pa_x) \psi_1^\da &=& -m\psi_2 -g\psi_2^\da\psi_2\psi_1^\da\,,\label{e3}\\
 \rmi(\pa_t + \pa_x) \psi_2^\da &=& -m\psi_1^\da - g\psi_1^\da\psi_1\psi_2^\da\,.\label{e4}
\end{eqnarray}
In components, the set of differential equations (\ref{e5.1}) and (\ref{e5.2}) read,
\bse\br
 \pa_x \Psi_1 &=& \left[\frac{\rmi g}{2}\rho_- +\frac{\rmi m}{2}\(\lm^2-\lm^{-2}\)\]\Psi_1 +q_1\Psi_3, \label{e5.8}\\
 \pa_x \Psi_2 &=& -\left[\frac{\rmi g}{2}\rho_- -\frac{\rmi m}{2}\(\lm^2-\lm^{-2}\)\]\Psi_2 +q_2\Psi_3, \label{e5.9}\\
 \pa_x \Psi_3 &=& r_1\Psi_1 + r_2 \Psi_2 +\rmi m \(\lm^2 -\lm^{-2}\) \Psi_3, \label{e5.10}
\er\ese
and
\bse\br
 \pa_t \Psi_1 &=& -\left[A-\frac{\rmi m}{2}\(\lm^2+\lm^{-2}\)\]\Psi_1 +B_1\Psi_3, \label{e5.11}\\
 \pa_t \Psi_2 &=& \left[A+\frac{\rmi m}{2}\(\lm^2+\lm^{-2}\)\]\Psi_2 +B_2\Psi_3, \label{e5.12}\\
 \pa_t \Psi_3 &=& C_1\Psi_1 +C_2\Psi_2 + \rmi m\(\lm^2 +\lm^{-2}\)\Psi_3. \label{e5.13}
\er\ese
Now, by defining the auxiliary functions $\G_{21} = \Psi_2\Psi_1^{-1}$ and $\G_{31}=\Psi_3\Psi_1^{-1}$, we obtain a first conservation equation from (\ref{e5.8}) and (\ref{e5.11}), namely,
\br
 \pa_t\[q_1\G_{31} +\frac{\rmi g}{2}\rho_-\] = \pa_x\[B_1\G_{31} -A\],
\er
where  $\G_{21}$ and $\G_{31}$ satisfy the following coupled Riccati equations for the $x$-part,
\bse\br
 \pa_x \G_{21} &=& -(\rmi g \rho_- )\G_{21} +q_2\G_{31} -q_1 \G_{21}\G_{31},\label{e4.66}\\
 \pa_x \G_{31} &=& r_1 +r_2\G_{21} - \frac{\rmi}{2}\left[g\rho_- -m\(\lm^2-\lm^{-2}\)\]\G_{31} ,\label{e4.67}
\er\ese
and for the $t$-part,
\bse\br
 \pa_t \G_{21} &=& 2A \G_{21} +B_2\G_{31} -B_1\G_{21}\G_{31},\\
 \pa_t \G_{31} &=& C_1 +C_2 \G_{21} +\left[A + \frac{\rmi m}{2}\(\lm^2+\lm^{-2}\)\]\G_{31}\,.
\er\ese
Now, by firstly considering an expansion in inverse powers of $\lm$ for the auxiliary functions as,
\br
 \G_{ij}(x,t;\lm) &=& \sum_{k=1}^{\infty} \frac{\G_{ij}^{(k)}(x,t)}{\lm^k},
\er
and inserting this in the Riccati equations (\ref{e4.66}) and (\ref{e4.67}) we find that the first coefficients of the expansion are given by
\bse\br
 \fl \G_{31}^{(1)} &=& \sqrt{\frac{2g}{m}}\,\psi_1^\da, \qquad \quad \G_{31}^{(2)} \,=\, -\(\sqrt{\frac{2g}{m}}\,\psi_1\) \,\G_{21}^{(1)}, \\
\fl  \G_{31}^{(3)} &=& -\frac{2\rmi}{m}\left[\sqrt{\frac{2g}{m}}\big(\pa_x\psi_1^\da\big) + \rmi\sqrt{\frac{mg}{2}}\(\psi_2^\da-\psi_1\G_{21}^{(2)}\) + \frac{\rmi g}{2}\sqrt{\frac{2g}{m}}(\psi_2^\da\psi_2)\psi_1^\da\],
\er\ese
where $\G_{21}^{(1)}$ and $\G_{21}^{(2)}$ satisfy the following differential equations,
\bse\br
 \pa_x \G_{21}^{(1)} &=& -\rmi g\(\psi_2^\da\psi_2+\psi_1^\da\psi_1\)\G_{21}^{(1)},\\
 \pa_x\G_{21}^{(2)} &=&   -\rmi g\(\psi_2^\da\psi_2+\psi_1^\da\psi_1\)\G_{21}^{(2)} +\frac{2g}{m} \(\psi_1^\da\pa_x\psi_1^\da\) +2\rmi g\,\psi_1^\da \psi_2^\da.
\er\ese
Then, we find out in this case that the associated generating function of the conserved quantities are given by,
\br
 {\tt I}_1 &=& \int_{-\infty}^{\infty} \rmd x \, \left[q_1\G_{31} +\frac{\rmi g}{2}\rho_- \],\label{e5.24.}
\er
and substituting the respective coefficients for the auxiliary function $\G_{31}$ in the expansion in $\lm$, we found that the lowest conserved quantities are given,
\bse\br 
 \fl {\tt I}_1^{(0)} &=& \frac{\rmi g}{2}\int_{-\infty}^{\infty} \rmd x \, \left[\psi_2^\da\psi_2+\psi_1^\da\psi_1\],\label{e5.25}\\
  \fl {\tt I}_{1}^{(2)} &=& \int_{-\infty}^{\infty} \rmd x \,\left[-\frac{2g}{m}\(\psi_1\pa_x\psi_1^\da\) + \rmi g \(\psi_2^\da\psi_1 +\psi_1^\da\psi_2\) +\frac{\rmi g^2}{m}\(\psi_2^\da\psi_2\psi_1^\da\psi_1\) \right].\label{e5.26}
\er\ese
%
Now, to compute a second infinite set of conserved quantities we have to expand the auxiliary functions around $\lm=0$, i.e, in positive powers of $\lm$, as follows
\br
 \G_{ij}(x,t;\lm) &=& \sum_{k=1}^{\infty} \what{\G}_{ij}^{(k)}(x,t) \,\lm^k .
\er
By following the same procedure, we obtain that the respective first few coefficients for each expansion are given by,
\bse\br
\fl \what{\G}_{31}^{(1)} &=&-\sqrt{\frac{2g}{m}}\psi_2^\da, \qquad \quad \what{\G}_{31}^{(2)} \,=\, -\(\sqrt{\frac{2g}{m}}\,\psi_2\)\what{\G}_{21}^{(1)}, \\
\fl \what{\G}_{31}^{(3)} &=& -\frac{2\rmi}{m}\left[\sqrt{\frac{2g}{m}}\big(\pa_x\psi_2^\da\big) - \rmi\sqrt{\frac{mg}{2}}\(\psi_1^\da+\psi_2\,\what{\G}_{21}^{(2)}\) -\frac{\rmi g}{2}\sqrt{\frac{2g}{m}}(\psi_1^\da\psi_1)\psi_2^\da\],
\er\ese
with
\bse\br
 \pa_x\what{\G}_{21}^{(1)} &=& \rmi g\(\psi_2^\da\psi_2+\psi_1^\da\psi_1\)\what{\G}_{21}^{(1)},\\
 \pa_x\what{\G}_{21}^{(2)} &=& \rmi g\(\psi_2^\da\psi_2+\psi_1^\da\psi_1\)\what{\G}_{21}^{(2)}-\frac{2g}{m} \(\psi_2^\da\pa_x\psi_2^\da\) -2\rmi g\,\psi_1^\da \psi_2^\da.
\er\ese
Then, we have that the corresponding first charges associated to this expansion of $\G_{31}$ are given as follows,
\bse\br
\fl \what{\tt{I}}_1^{(0)} &=& -\frac{\rmi g}{2}\int_{-\infty}^{\infty} \rmd x \, \left[\psi_2^\da\psi_2+\psi_1^\da\psi_1\],\label{e5.34}\\
\fl  \what{\tt{I}}_1^{(2)}&=& \int_{-\infty}^{\infty} \rmd x \left[-\frac{2g}{m}\(\psi_2\pa_x\psi_2^\da\)- \rmi g \(\psi_2^\da\psi_1 +\psi_1^\da\psi_2\) -\frac{\rmi g^2}{m}\(\psi_2^\da\psi_2\psi_1^\da\psi_1\) \right]\!\!. \qquad \mbox{}\label{e5.35}
\er\ese
Clearly, these charges are not totally real and therefore it is necessary to add the hermitian conjugate terms. To do that, we need to consider other contributions coming from two more conservation equations that can be derived using (\ref{e5.9}), (\ref{e5.10}), (\ref{e5.12}) and (\ref{e5.13}), namely
\bse\br
 \pa_t\[q_2\G_{32} -\frac{\rmi g}{2}\rho_-\] &=& \pa_x\bigg[B_2\G_{32} + A\bigg],\label{e5.36}\\
 \pa_t\bigg[r_1\G_{13} + r_2 \G_{23} \bigg] & =& \pa_x\bigg[C_1\G_{13} +C_2 \G_{23}\bigg],\label{e5.37}
\er\ese
where we have introduced some other auxiliary functions $\G_{12} =\Psi_1\Psi_2^{-1}$, $\G_{32} =\Psi_3\Psi_2^{-1}$, $\G_{13} =\Psi_1\Psi_3^{-1}$, and $\G_{23} =\Psi_2\Psi_3^{-1}$. It is very easy to check that the set of Riccati equations satisfied by these auxiliary functions can be written as,
\bse\br
 \pa_x \G_{12}  &=& \rmi g\,\rho_- \G_{12} +q_1\G_{32} -q_2\G_{12}\G_{32}, \\
 \pa_x \G_{32} &=& r_2 +r_1\G_{12} +\frac{\rmi}{2}\(g\rho_- + m(\lm^2 - \lm^{-2})\)\G_{32},\\
 \pa_x \G_{13} &=& q_1 +\frac{\rmi}{2}\left[g\rho_- - m\(\lm^2-\lm^{-2}\) \] \G_{13} +r_2 \G_{13}\G_{23},\\
 \pa_x \G_{23} &=& q_2 -\frac{\rmi}{2}\left[g\rho_- +m \(\lm^{2} - \lm^{-2}\) \right]\G_{23} -r_1\G_{13}\G_{23},
\er\ese
and
\bse\br
 \pa_t \G_{12}  &=& -2A\,\G_{12} +B_1\G_{32} +B_2\G_{12}\G_{32},\\
 \pa_t \G_{32} &=& C_2 +C_1 \G_{12} - \left[A-\frac{\rmi m}{2}\(\lm^2+\lm^{-2}\)\]\G_{32}\\
 \pa_t \G_{13} &=& B_1 -\left[A+\frac{\rmi m}{2}\(\lm^2+\lm^{-2}\)\]\G_{13} +C_2\G_{13}\G_{23}\\
 \pa_t \G_{23} &=& B_2 +\left[A-\frac{\rmi m}{2}\(\lm^2+\lm^{-2}\)\]\G_{23} -C_1\G_{13}\G_{23}.
\er\ese
Now, these equations are solved by expanding each of the auxiliary functions in positive and negative powers of the spectral parameter $\lm$. Performing similar computations, the first few coefficients for these auxiliary functions can be determined, and the results read
\bse\br
 \fl \G_{23}^{(1)} & =&\sqrt{\frac{2g}{m}}\,\psi_1^\da, \qquad \G_{23}^{(1)} \,=\, \G_{32}^{(1)}  \,=\, \sqrt{\frac{2g}{m}}\psi_1 , \qquad {\what\G}_{13}^{(1)} \,=\,- {\what\G}_{32}^{(1)}\,=\,\sqrt{\frac{2g}{m}}\,\psi_2, \\
\fl {\what\G}_{23}^{(1)} &=& \sqrt{\frac{2g}{m}}\,\psi_2^\da, \qquad  \G_{32}^{(2)} =\sqrt{\frac{2g}{m}}\,\psi_1^\da,\qquad  {\what \G}_{32}^{(2)}\,=\, -\sqrt{\frac{2g}{m}}\,\psi_2^\da ,\\
\fl \G_{13}^{(2)} &=& \G_{23}^{(2)} \,=\, {\hat\G}_{13}^{(2)}\,=\, \hat{\G}_{23}^{(2)} \,=\,0, \qquad 
\er\ese
and
\bse\br
\fl \G_{32}^{(3)} &=& \frac{2}{m}\left[\rmi\sqrt{\frac{2g}{m}}\,(\pa_x\psi_1) + \sqrt{\frac{mg}{2}}\(\psi_2 +\psi_1^\da\, \G_{122}\) +\frac{g}{2}\sqrt{\frac{2g}{m}}\(\psi_2^\da\psi_2\)\psi_1 \right] ,\\
  \fl  {\what\G}_{32}^{(3)} &=&\frac{2}{m}\left[- \rmi\sqrt{\frac{2g}{m}}\,(\pa_x\psi_2) + \sqrt{\frac{mg}{2}}\(\psi_1 -\psi_2^\da\, {\what\G}_{122}\) +\frac{g}{2}\sqrt{\frac{2g}{m}}\(\psi_1^\da\psi_1\)\psi_2 \right], \\
 \fl \G_{13}^{(3)} &=& \frac{2}{m}\left[ -\rmi\sqrt{\frac{2g}{m}}\,(\pa_x\psi_1) - \sqrt{\frac{mg}{2}}\,\psi_2 -\frac{g}{2}\sqrt{\frac{2g}{m}}\(\psi_2^\da\psi_2\)\psi_1 \right], \\
\fl {\what\G}_{13}^{(3)} &=& \frac{2}{m}\left[ -\rmi\sqrt{\frac{2g}{m}}\,(\pa_x\psi_2) + \sqrt{\frac{mg}{2}}\,\psi_1 +\frac{g}{2}\sqrt{\frac{2g}{m}}\(\psi_1^\da\psi_1\)\psi_2 \right], \\
\fl  \G_{23}^{(3)} &=&\frac{2}{m}\left[ \rmi\sqrt{\frac{2g}{m}}\,(\pa_x\psi_1^\da) - \sqrt{\frac{mg}{2}}\,\psi_2^\da -\frac{g}{2}\sqrt{\frac{2g}{m}}\(\psi_2^\da\psi_2\)\psi_1^\da \right], \\
\fl {\what\G}_{23}^{(3)} &=&\frac{2}{m}\left[ -\rmi\sqrt{\frac{2g}{m}}\,(\pa_x\psi_2^\da) - \sqrt{\frac{mg}{2}}\,\psi_1^\da -\frac{g}{2}\sqrt{\frac{2g}{m}}\(\psi_1^\da\psi_1\)\psi_2^\da \right],
\er\ese
together with the following relations,
\bse\br
 \pa_x \G_{12}^{(1)} &=& \rmi g \(\psi_1^\da\psi_1 +\psi_2^\da\psi_2\)\G_{12}^{(1)}, \\
 \pa_x {\what\G}_{12}^{(1)} &=& -\rmi g \(\psi_1^\da\psi_1 +\psi_2^\da\psi_2\){\what\G}_{12}^{(1)}, \\
 \pa_x \G_{12}^{(2)} &=& \rmi g \(\psi_1^\da\psi_1 +\psi_2^\da\psi_2\)\G_{12}^{(2)} +\frac{2g}{m} \(\psi_1\pa_x\psi_1\) -2\rmi g\(\psi_1\psi_2\), \\
 \pa_x {\what\G}_{12}^{(2)} &=& -\rmi g \(\psi_1^\da\psi_1 +\psi_2^\da\psi_2\){\what\G}_{12}^{(2)} -\frac{2g}{m} \(\psi_2\pa_x\psi_2\) + 2\rmi g\(\psi_1\psi_2\).
\er\ese
Now, we will compute the corresponding conserved quantities from the conservation equations (\ref{e5.36}) and (\ref{e5.37}), namely
\bse\br
 {\tt I}_2 &=& \int_{-\infty}^{\infty}\rmd x \left[q_2 \G_{32} -\frac{\rmi g}{2}\rho_- \right], \label{e5.50}\\[0.1cm]
 {\tt I}_3 &=&\int_{-\infty}^{\infty}\rmd x \,\bigg[r_1 \G_{13} + r_2\G_{23}\bigg].\label{e5.51}
\er\ese
Therefore, by a straightforward substitution of the each expansion coefficient, we easily get the following results,
{\small
\bse\br 
 \fl {\tt I}_2^{(0)} &=& -\frac{\rmi g}{2}\int_{-\infty}^{\infty} \rmd x \, \left[\psi_2^\da\psi_2+\psi_1^\da\psi_1\] \,\,=\,\,  - \,{\what{\tt I}}_2^{(0)}, \,\qquad{\tt I}_3^{(0)} = 0 \,=\, {\what{\tt I}}_3^{(0)}, \\
\fl  {\tt I}_2^{(2)} &=& \int_{-\infty}^{\infty} \rmd x \,\left[-\frac{2g}{m}\(\psi_1^\da\pa_x\psi_1\) + \rmi g \(\psi_2^\da\psi_1 +\psi_1^\da\psi_2\) +\frac{\rmi g^2}{m}\(\psi_2^\da\psi_2\psi_1^\da\psi_1\) \right], \\
\fl {\what{\tt I}}_2^{(2)} &=& \int_{-\infty}^{\infty} \rmd x \,\left[-\frac{2g}{m}\(\psi_2^\da\pa_x\psi_2\) - \rmi g \(\psi_2^\da\psi_1 +\psi_1^\da\psi_2\) -\frac{\rmi g^2}{m}\(\psi_2^\da\psi_2\psi_1^\da\psi_1\) \right], \\
\fl {\tt I}_3^{(2)} &=&\!\!\! \int_{-\infty}^{\infty}\rmd x \,\left[-\frac{2g}{m}\(\psi_1^\da\pa_x\psi_1+ \psi_1\pa_x\psi_1^\da\) +2 \rmi g \(\psi_2^\da\psi_1 +\psi_1^\da\psi_2\) +\frac{2\rmi g^2}{m}\(\psi_2^\da\psi_2\psi_1^\da\psi_1\) \right]\\
\fl {\what{\tt I}}_3^{(2)}&=&\!\!\! \int_{-\infty}^{\infty} \rmd x \,\left[-\frac{2g}{m}\(\psi_2^\da\pa_x\psi_2+ \psi_2\pa_x\psi_2^\da\) -2\rmi g \(\psi_2^\da\psi_1 +\psi_1^\da\psi_2\) -\frac{2\rmi g^2}{m}\(\psi_2^\da\psi_2\psi_1^\da\psi_1\) \right].
\er\ese}
\normalsize Then, from all these conserved quantities together with the ones derived in (\ref{e5.25}), (\ref{e5.26}), (\ref{e5.34}), and (\ref{e5.35}), we can notice that
\br
  {\tt I}_1^{(n)} + {\tt I}_2^{(n)}&=& {\tt I}_3^{(n)}, \quad
 {\what{\tt I}}_1^{(n)} + {\what{\tt I}}_2^{(n)} \,=\, {\what{\tt I}}_3^{(n)}.
\er
Therefore, it is convenient to define the following quantities,
\bse\br
  {\bb I}^{(0)} &=& ({\tt I}_1^{(0)} - {\tt I}_2^{(0)} -{\tt I}_3^{(0)} ), \qquad  {\what{\bb I}}^{(0)} \,=\, \big(\,{\what{\tt I}}_1^{(0)} - {\what{\tt I}}_2^{(0)}- {\what{\tt I}}_3^{(0)}\big),\\
 {\bb I}^{(2)} &=& ({\tt I}_1^{(2)} + { \tt I}_2^{(2)} +{ \tt I}_3^{(2)}), \qquad {\what{\bb I}}^{(2)} \,=\, \big({\what{\tt I}}_1^{(2)} +{\what{\tt I}}_2^{(2)} + {\what{\tt I}}_3^{(2)}\big),
\er\ese
in order to get the usual conserved number of occupation, energy and momentum for the GT model by performing a simple combination, namely
\bse\br
\fl N &=& \frac{1}{2\rmi g}\left[{\bb I}^{(0)}-{\what{\bb I}}^{(0)}\] \,=\, \int_{-\infty}^{\infty} \rmd x \, \left[\psi_2^\da\psi_2+\psi_1^\da\psi_1\],\\[0.1cm]
\fl E &=& \frac{m}{8\rmi g}\left[{\bb I}^{(2)}-{\what{\bb I}}^{(2)}\] \,=\, \int_{-\infty}^{\infty} \rmd x \,\left[\frac{\rmi}{2}\bigg(\psi_1\pa_x\psi_1^\da + \psi_1^\da\pa_x\psi_1-\psi_2\pa_x\psi_2^\da - \psi_2^\da\pa_x\psi_2\bigg) \right.\nonumber \\ 
\fl &\mbox{}&\left. \qquad \qquad \qquad\qquad \qquad \quad\,\, +\,m\,\big(\psi_2^\da\psi_1+\psi_1^\da\psi_2 \big) +g \,\psi_2^\da\psi_2\psi_1^\da\psi_1\],\qquad \qquad \qquad \mbox{}\\[0.1cm]
\fl P &=&  \frac{m}{8\rmi g}\left[{\bb I}^{(2)} +{\what{\bb I}}^{(2)}\] \,=\,\int_{-\infty}^{\infty} \rmd x \,\left[\frac{\rmi}{2}\bigg(\psi_1\pa_x\psi_1^\da + \psi_1^\da\pa_x\psi_1+\psi_2\pa_x\psi_2^\da + \psi_2^\da\pa_x\psi_2\bigg)\]. 
\er\ese
In the next section, we show how this framework can be used in order to compute the modified conserved quantities from the defect matrix for the GT model.


\subsection{Modified integrals of motion from the defect matrix}
Let us implement a defect placed at the origin $x=0$, and the relation between the respective auxiliary wavefunctions $\Psit$ in $x<0$, and the corresponding $\Psi$ for $x>0$, is given by
\br
 \Psit(x,t;\lm) &=& K_{\rm G}(x,t;\lm) \,\Psi(x,t;\lm),
\er
where the defect matrix $K_G$ satisfied the gauge transformations,
\bse\br
 \pa_t K_{\rm G} &=& \widetilde{V}\,K_{\rm G} -K_{\rm G}\, V, \\
 \pa_x K_{\rm G} &=& \widetilde{U}\, K_{\rm G} - K_{\rm G}\, U.
\er\ese
The explicit form of the defect matrix $K_{\rm G}$ was also computed in \cite{Ale2} and can be written in the following simple form,
\br
 K_{\rm G} &=& \left[
   \begin{array}{ccccc} K_- &\mbox{}& 0 &\mbox{}& \sqrt{\frac{2g}{m}}X\\[0.3cm]
   0 & \mbox{}& K_+  & \mbox{}& -\sqrt{\frac{2g}{m}}X^\da \\[0.3cm] 
   \sqrt{\frac{2g}{m}}X^\da & \mbox{\qquad}&-\sqrt{\frac{2g}{m}}X  & \mbox{\qquad}&  -(\lm + \rmi (\lm a)^{-1}) 
   \end{array} 
 \right] \label{e5.68}
\er
where the elements $K_\pm$ are given by, 
\br
 K_{\pm} &=& \lm\exp\[\pm\frac{\rmi ga}{2m}X^\da X\] -\rmi (\lm a)^{-1}\exp\[\mp\frac{\rmi ga}{2m}X^\da X\]\nonumber\\
 &=& \left[\lm-\rmi (\lm a)^{-1}\] \pm \frac{\rmi ga}{2m}\left[\lm+ \rmi (\lm a)^{-1}\]X^\da X, 
\er
and the Grassmannian boundary fields $X$ and $X^\da$ satisfy the following defect conditions, 
\begin{eqnarray}
\fl X &=& (\psit_1 + \psi_1 ) + {{i ag}\over {2m}}\psit_1  X^{\dagger} X 
 \,\,=\,\, i a^{-1}(\psi_2-\psit_2) -\frac{g}{2m}X^\da X \psi_2, \nonumber \\
\fl X^\da &=& (\psit_1^{\da} + \psi_1^{\da	} ) - {{\rmi ag}\over {2m}}\psit_1^{\da}  X^{\dagger} X \,\,\,\,=\,\, -\rmi a^{-1}(\psi_2^{\da}-\psit_2^{\da}) -\frac{g}{2m}X^\da X \psi_2^{\da}, \qquad \mbox{} \label{e5.71}
\end{eqnarray}
together with their respective time derivatives,
\bse\begin{eqnarray}
\fl \pa_t X &=& {{m}\over {2a}} (\psi_1- \psit_1) - {{\rmi m}\over {2}} (\psi_2 + \psit_2) -\frac{\rmi g}{4}\left[\psit_1^{\da}\psit_1 + \psi_1^{\da}\psi_1 + \psit_2^{\da}\psit_2 + \psi_2^{\da}\psi_2\right] X, \\
\fl \pa_t X^{\dagger}  &=& {{m}\over {2a}} (\psi_1^{\da} - \psit_1^{ \da }) + {{\rmi m}\over {2}} (\psi_2^{ \da } + \psit_2^{ \da })+\frac{\rmi g}{4}\left[\psit_1^{\da}\psit_1+ \psi_1^{\da}\psi_1  + \psit_2^{\da}\psit_2 + \psi_2^{\da}\psi_2\right] X^\da , \label{eqn4.13}
\end{eqnarray}\ese
and their $x$-derivatives,
\bse\br
\fl \pa_x X &=& {{m}\over {2a}} (\psi_1 - \psit_1) +{{\rmi m}\over {2}} (\psi_2 + \psi_2) -\frac{\rmi g}{4}\left[\psit_1^{\da}\psit_1 + \psi_1^{\da}\psi_1  - \psit_2^{\da}\psit_2 - \psi_2^{\da}\psi_2 \right] X, \\
\fl \pa_x X^{\dagger}  &=& {{m}\over {2a}} (\psi_1^{\da } - \psit_1^{\da}) - {{\rmi m}\over {2}} (\psi_2^{\da} + \psit_2^{\da})+\frac{\rmi g}{4}\left[\psit_1^{\da}\psit_1 + \psi_1^{\da}\psi_1  - \psit_2^{\da}\psit_2 - \psi_2^{\da}\psi_2\right] X^\da , \label{e5.75.}
\er\ese
which correspond precisely to  the auto-B\"acklund transformations for the classical GT model \cite{Ale,Ize}. 

Let us consider now the defect contributions to the conserved quantities. As we have discussed across this work, the entries of the defect matrix determine the modified conserved quantities from (\ref{formu}). First of all, let us consider the first set of conserved quantities given by (\ref{e5.24.}) in the presence of  a defect,
\br
 {\cal I}_1 &=& \int_{-\infty}^0 \rmd x \left[\widetilde{q}_1\widetilde{\G}_{31} +\frac{\rmi g}{2}\widetilde{\rho}_- \] + \int_{0}^{\infty} dx \left[q_1\G_{31} +\frac{\rmi g}{2}\rho_- \].
\er
Then, taking the time-derivative and using the formula we found that ${\cal I}_1 + {D}_1$ is conserved, where the defect contribution $D_1$ to this first set of conserved quantities is explicitly given by
\br
 D_1 &=& - \ln \bigg[K_{11} + K_{12} \G_{21} +K_{13}\G_{31}\bigg]\bigg|_{x=0}.
\er
Hence, by taking the both expansions in negative and positive powers of $\lm$ and the explicit form of the defect matrix (\ref{e5.68}), we get
\bse\br
\fl {D}_1^{(0)} &=& \(\frac{\rmi ga}{2m}\)X^\da X, \qquad \qquad \quad\, {\what{D}}_1^{(0)} \,=\, -\(\frac{\rmi ga}{2m}\)X^\da X,  \\[0.1cm]
 \fl  {D}_1^{(2)} &=&  -\frac{g}{m}X^\da X-\frac{2g}{m}X\psi_1^{\da}, \qquad 
{\what{D}}_1^{(2)} = -\frac{ga^2}{m}X^\da X + \frac{2\rmi ag}{m}X\psi_2^{\da}.
\er\ese
In a similar way, repeating the computations for the other two generating functions (\ref{e5.50}) and (\ref{e5.51}), we find that the respective defect contributions are given by,
\bse\br
 D_2 = -\ln \left[K_{21}\G_{12} +K_{22} +K_{23}\G_{32}\]\bigg|_{x=0}, \\  
 D_3 = -\ln \left[K_{31}\G_{13} + K_{32}\G_{23} + K_{33}\]\bigg|_{x=0}.
\er\ese
Using them, we obtain
\bse\br
\fl {D}_2^{(0)} &=& -\(\frac{\rmi ga}{2m}\)X^\da X, \qquad \, {\what{D}}_2^{(0)} \,=\, 
\(\frac{\rmi ga}{2m}\)X^\da X, \qquad  {D}_3^{(0)} \,=\, 0 \,=\, {\what{D}}_3^{(0)}\,, \\
\fl {D}_2^{(2)} &=& \frac{g}{m}X^\da X- \frac{2g}{m} X^\da\psi_1, \qquad \quad\,\,\,
\,{\what{D}}_2^{(2)} \,=\,\frac{ga^2}{m}X^\da X - \frac{2\rmi ag}{m} X^\da\psi_2,\quad \qquad \mbox{}\\
\fl {D}_3^{(2)} &=& - \frac{2g}{m} X^\da\psi_1 -\frac{2g}{m}X\psi_1^{\da},\qquad\,\,\, {\what{D}}_3^{(2)}\,=\,- \frac{2\rmi ag}{m} X^\da\psi_2 +\frac{2\rmi ag}{m}X\psi_2^{\da}.\qquad \,\,\, \mbox{}
\er\ese
As it was expected, we also have the relations ${{D}}_3^{(n)} = {D}_1^{(n)}  + {D}_2^{(n)}$ and ${\what{D}}_3^{(n)} = {\what{D}}_1^{(n)}  +  {\what{D}}_2^{(n)}$. Then, defining by analogy the following defect quantities,
\bse\br
 {\cal D}^{(0)} &=&  {D}_1^{(0)}  -  {D}_2^{(0)} -{D}_3^{(0)} \,\,=\,\, \(\frac{\rmi ga}{m}\)X^\da X, \\
  {\what{\cal D}}^{(0)}  &=&  {\what{D}_1}^{(0)}  -  {\what{D}}_2^{(0)} - {\what{D}}_3^{(0)} \,\,=\,\,- \(\frac{\rmi ga}{m}\)X^\da X,\\
 {\cal D}^{(2)} &=& { D}_1^{(2)}  +{D}_2^{(2)}+ D_3^{(2)}\,\,=\,\,- \frac{4g}{m} X^\da\psi_1 -\frac{4g}{m}X\psi_1^{\da}, \\
{\what{\cal D}}^{(2)} &=& {\what D}_1^{(2)}  + {\what D}_2^{(2)} +  {\what{D}}_3^{(2)}\,\,=\,\,- \frac{4\rmi ag}{m} X^\da\psi_2 +\frac{4\rmi ag}{m}X\psi_2^{\da},
\er\ese
the corresponding defect number of occupation, energy and momentum can be written in the following way,
\bse\br
 \fl N_D &=& \frac{1}{2\rmi g}\({\cal D}^{(0)} -  {\what{\cal D}}^{(0)} \) \,\,=\,\, \frac{a}{m}X^\da X,\label{e5.93..} \\
\fl  E_D &=& \frac{m}{8\rmi g}\({\cal D}^{(2)}  -{\what{\cal D}}^{(2)}\) \,\,=\,\, \frac{\rmi}{2}\left[(X^\da\psi_1 + X\psi_1^{\da}) -\rmi a(X^\da\psi_2 - X\psi_2^{\da}) \right], \quad \mbox{}\\
\fl   P_D &=& \frac{m}{8\rmi g}\({\cal D}^{(2)}  +{\what{\cal D}}^{(2)}\) \,\,=\,\, \frac{\rmi}{2}\left[(X^\da\psi_1+X\psi_1^{\da}) + \rmi a(X^\da\psi_2 - X\psi_2^{\da}) \right].
\er\ese
Notice that, we can rewrite these results by using the B\"acklund transformations (\ref{e5.70.}) and (\ref{e5.71}) in a more convenient form, by eliminating the auxiliary fields $X$ and $X^\da$. The results are the following, 
\br
\fl 
 E_D &=& \frac{\rmi}{2}\left[\psit_1^{\da}\psi_1-\psi_1^{\da}\psit_1 +\psit_2^{\da}\psi_2 -\psi_2^{\da}\psit_2 \right] -\frac{ag}{2m}(\psit_1^{\da}\psit_1\psi_1^{\da}\psi_1)-\frac{g}{2ma}(\psit_2^{\da}\psit_2\psi_2^{\da}\psi_2),\label{e5.90}\\[0.2cm]
\fl  P_D &=& \frac{\rmi}{2}\left[\psit_1^{\da}\psi_1-\psi_1^{\da}\psit_1 -\psit_2^{\da}\psi_2 +\psi_2^{\da}\psit_2 \right] -\frac{ag}{2m}(\psit_1^{\da}\psit_1\psi_1^{\da}\psi_1)+\frac{g}{2ma}(\psit_2^{\da}\psit_2\psi_2^{\da}\psi_2).\label{e5.91}
\er
Then, we have particularly derived in an alternative way the defect energy and momentum for the Grassmannian Thirring model in the presence of type-II defects. These results are in complete agreement with the ones obtained based on variational principles \cite{Ale}. The advantage of the approach used in this work is that it provides us a formal way to compute explicitly  an infinite number of conserved quantities ensuring integrability of these models. However, taking into account recent developments in the area of integrable defects \cite{Anastasia} it turns out interesting to investigate how to provide a Hamiltonian formulation to include the models discussed in this work since the role of the degree of freedom corresponding to the defect itself needs to be clarified. In the next section, we discuss some useful ideas in addressing this question but a more complete description of the Hamiltonian approach will be postponed for a future work.

\section{Comments on Liouville integrability}

So far, an infinite set of independent modified conserved quantities arising from the defect contributions have been systematically constructed through a general formula derived from a variant of the classical inverse scattering method, which are from our point of view sufficient for these kind of defects to be regarded as integrable. However, the question of the involutivity of such quantities (required to discuss complete integrability in the sense of Liouville)  still has to be answered. In this section we address some relevant facts related with  this problem.

Certainly, the Hamiltonian formulation of the classical inverse scattering method, which is essentially based on the concept of a classical $r$-matrix \cite{Skly2}, is perhaps the most elegant and convenient framework to discuss involutivity. Let us start with the main aspects of the method in order to discuss this issue in the bulk. In the inverse scattering method the construction of the action-angle variables depends basically on the entries of the monodromy matrix $\tau(\lm) = T(\infty,-\infty;\lm)$, where
\br
 T(x,y;\lm) = {\cal P} \exp \left\{\int_y^x U(z;\lm) dz\right\},
\er
is the transition matrix, $U(x;\lm)$ is the $x$-part of the Lax (\ref{s2e0.2}) at a given time, and ${\cal P}$ being the path ordering. 
As it was noticed in \cite{Skly2,Sky}, the existence of the classical $r$-matrix, an $m^2\times m^2$ matrix which satisfies the relation
\br
 \fl \poisson{U(x;\lm_1)}{U(y;\lm_2)} = \d(x-y)\Big[r(\lm_1,\lm_2), \, U(x;\lm_1) \otimes {\textsc I}_2 +  {\textsc I}_2\otimes U(y;\lm_2)  \Big],
\er
permits us to write down the Poisson brackets between matrix elements of the transition matrix in the following form,
\br
\poisson{T(x,y;\lm_1)}{T(x,y;\lm_2)} = \Big[r(\lm_1,\lm_2), \, T(x,y;\lm_1) \otimes T(x,y;\lm_2)  \Big],\label{e6.3}
\er
from which it is derived that logarithm of the traces of the monodromy matrix commute for different values of the spectral parameter, namely
\br
 \left\{\ln\tau(\lm_1),\ln\tau(\lm_2)\right\} = 0.\label{e6.4}
\er
Expanding (\ref{e6.4}) with respect to $\lm_1$ and $\lm_2$, we get the involutivity of the conserved quantities $\{I_n\}$, which means that $\t(\lm)$ is the generating functional for the integrals of motion. Let us mention that the explicit form of the $r$-matrix for the massive Thirring models can be found in \cite{Skly2}.

Now, let us discuss how the classical $r$-matrix approach is modified by including jump-defect (or point like-defect) in the system. As it was noticed in \cite{Hab2} and more recently in \cite{Anastasia}, the description of an integrable defect in the $r$-matrix approach requires to introduce a modified transition matrix,
\br
 {\cal T}(x,y;\lm) = {T}(x,0^{+};\lm) \,K^{-1}(0;\lm)\,{\widetilde T}(0^-,y;\lm),\label{e6.5}
\er
which is a combined bulk-defect transition matrix, where $ {T}(x,0^{+};\lm)$ and ${\widetilde T}(0^-,y;\lm)$ are the bulk transition matrices corresponding to $x>0$ and $x<0$ respectively, and $K(\lm)\equiv K(0;\lm)$ is the defect matrix whose entries are evaluated in the single point $x=0$. The key point in order to show Liouville integrability is to require that the defect matrix satisfy the Poisson algebra (\ref{e6.3}), namely,
\br
  \poisson{K^{-1}(\lm_1)}{K^{-1}(\lm_2)} =\Big[r(\lm_1,\lm_2), \, K^{-1}(\lm_1) \otimes K^{-1}(\lm_2)\Big],
\er
where $r(\lm_1,\lm_2)$ is the same classical $r$-matrix for the bulk transition matrices. Hence, the above requirement is a sufficient condition to obtain the important result,
\br
\poisson{{\cal T}(x,y;\lm_1)}{{\cal T}(x,y;\lm_2)} = \Big[r(\lm_1,\lm_2), \, {\cal T}(x,y;\lm_1) \otimes {\cal T}(x,y;\lm_2)  \Big],\label{e6.3}
\er
which guarantees the existence of the infinite set of modified conserved quantities. Similar to the bulk theory, the explicit form of these integrals of motion can be extracted by introducing the following representation for the bulk transition matrix \cite{Tak},
\br
 T(x,y;\lm) = \(1+W(x;\lm)\) \,e^{Z(x,y;\lm)}\,\(1+W(y;\lm)\)^{-1},
\er
where $W(x;\lm)$ is an off-diagonal and $Z(x,y;\lm)$ a diagonal matrix. Then, the logarithm of the trace of the modified monodromy matrix (\ref{e6.5}) is the generating function of the modified conserved quantities, where the defect contributions in an appropriate expansion in $\lm$, read \cite{Hab2,Anastasia}: 
\br
 D(\lm) = \ln\left[\Big(1+W(0^+,\lm)\Big)^{-1}\,K^{-1}(\lm) \, \Big(1+{\widetilde W}(0^-,\lm)\Big) \right]_{ii},
\er
where the subscript $ii$ denotes the leading term coming from the trace of the modified monodromy matrix for the given expansion. At first sight, it seems not to exist a direct relationship between the above result and the generating function (\ref{formu}) what we have worked with. However, note that $W(x;\lm)$ satisfy a matrix Riccati equation similar to (\ref{equa2.4}), which permits us to derive recursively its coefficients in an asymptotic series expansion as $\lm\to \infty$ and $\lm\to 0$, and to demonstrate order by order that the results are complete equivalent for the massive Thirring models. This analysis deserves more attention than we could give at this moment and is left for future investigations.

It is remarkable that the approach we have adopted (see also \cite{Ale3}), uses essentially an on-shell defect matrix which implies that its entries have non-vanishing Poisson brackets with the bulk monodromy matrices elements. This fact has already been outlined in \cite{zambon1} for the Hamiltonian formulation of the type-II defects in the sine-Gordon and Tzitz\'eica  model, where the defect conditions appear as a set of second class constraints on the fields, which induces a slight modification of the canonical Poisson brackets. This issue indeed can be solved by working firstly with the off-shell defect matrix to compute the Poisson brackets and then derive the constraints as consistency conditions in constructing the time-like operator in the Lax pair such that the zero curvature condition provides the same equations of motion as the ones coming from the Hamiltonian evolution derived via Poisson brackets \cite{Anastasia}.


\section{Concluding remarks and perspectives}
\label{sec:discussions}

In this paper, we have presented a systematic approach to the integrability problem of the Bosonic and Grassmannian Thirring models with type-II defects using the inverse scattering formalism. Such a formulation allows to compute explicitly the modified conserved quantities to all orders in terms of the defect matrix. 

We have followed the approach to defects in classical integrable field theories given in \cite{Cau}, to present for the $m\times m$ linear problem, a direct generalization  for the generating function of the defect contributions for the modified conserved quantities to all orders. We have successfully applied this procedure in the case of Bosonic ($m=2$), and Grassmannian ($m=3$) Thirring models. For the latter, we have recovered previous results obtained from the Lagrangian approach \cite{Ale}. 
For the Bosonic Thirring model, we have also derived explicitly the defect contributions for the energy and momentum. These results seem not to be reported elsewhere in the literature, and should be of crucial importance for  further studies on its possible Lagrangian description.

However, a remarkable aspect is that despite the duality relation between the sine-Gordon and Thirring models, it is not clear for the author why they allow different types of integrable defects, in the sense that the sine-Gordon model allows type-I and type-II defects, but the Thirring model apparently only allows type-II integrable defects. It is worth exploring in more detail the relation between these type-II defects through bosonization techniques in future developments.

Finally, taking into account the reasons already mentioned in section 6 about 
the classical $r$-matrix description of Liouville-integrable point-like defects in integrable field theories,
it should be interesting to investigate the aspects of the complete integrability of the Bosonic and Grassmannian Thirring models with type-II defects within the Hamiltonian framework. Perhaps the most important motivation to perform that further study is the possibility of considering the quantization by the so-called quantum inverse scattering method \cite{Sky,Fad,Kor}.  Some of these questions are expected to be developed in future investigations.


\ack
I am grateful to Professors Abraham H. Zimerman and Jos\'e F. Gomes for many valuable discussions, suggestions and encouraging me in doing this work. I also thank the referees for helpful suggestions and comments. I would like to thank to David Schmidtt for fruitful discussions, and Leandro H. Ymai for useful comments in the early stage of this work. I thank Professor Vincent Caudrelier for helpful comments on a draft of this paper and for letting me know the reference \cite{Tsu1} for multicomponent systems. Special thanks to my wife Suzana Moreira for reading the manuscript and helping me to improve my English. I would also like to thank Ag\^encia FAPESP S\~ao Paulo Research Foundation for financial support under the Ph.D scholarship \mbox{2008/06555-6.}


\appendix

\section{A brief review of Type-I defect sine-Gordon model}

The sine-Gordon equation
\br
 \pa_t^2 \vp -\pa_x^2 \vp = -m^2 \sin \vp, \label{esine}
\er
can be derived as a compatibility condition for the associated linear problem given by the Lax pair
\br
  U &=&
 \left[
   \begin{array}{cc} -\frac{\rmi}{4}\big( \pa_t \vp\big)   & q(\lm)\\[0.3cm]
r(\lm)  &   \frac{\rmi}{4}\big(\pa_t \vp\big)
   \end{array}
 \], \qquad 
 V \,=\,
 \left[
   \begin{array}{cc} -\frac{\rmi}{4}\big( \pa_x \vp\big)   & A(\lm) \\[0.3cm]
   B(\lm) & \frac{\rmi}{4}\big( \pa_x \vp\big) 
   \end{array}
 \],\qquad \mbox{}\label{laxsine}
\er
where it has been defined the fields, 
\br
q(\lm)&=& -\frac{m}{4}\big(\lm e^{\frac{\rmi\vp}{2}} -\lm^{-1}e^{-\frac{\rmi\vp}{2}} \big) , \qquad 
r(\lm) \,=\,     \frac{m}{4}\big(\lm e^{-\frac{\rmi\vp}{2}} -\lm^{-1}e^{\frac{\rmi\vp}{2}} \big),\\
A(\lm) &=& -\frac{m}{4}\big(\lm e^{\frac{\rmi\vp}{2}} +\lm^{-1}e^{-\frac{\rmi\vp}{2}} \big) , \qquad 
B(\lm) \,=\, \frac{m}{4}\big(\lm e^{-\frac{\rmi\vp}{2}} +\lm^{-1}e^{\frac{\rmi\vp}{2}} \big) .
\er
From the linear system, we can derive two conservation equations, namely
\br
 \pa_t\left[q\G_{21} -\frac{\rmi}{4}\(\pa_t\vp\)\right] = \pa_x\left[A\G_{21} -\frac{\rmi}{4}\(\pa_x\vp\) \right] \label{e6.7},\\
 \pa_t\left[r\G_{12} +\frac{\rmi}{4}\(\pa_t\vp\)\right] = \pa_x\left[B\G_{12} +\frac{\rmi}{4}\(\pa_x\vp\) \right] \label{e6.8},
\er
where the auxiliary functions $\G_{21}=\P_2\P_1^{-1}$ and $\G_{12} =\P_1\P_2^{-1}$ has been introduced, which satisfy the Riccati equations,
\br
\fl \pa_x\G_{21} &=& r +\frac{\rmi}{2}\(\pa_t\vp\) \G_{21} - q\G_{21}^2, \qquad \quad
  \pa_x\G_{12}\,=\,q -\frac{\rmi}{2}\(\pa_t\vp\) \G_{12} - r\G_{12}^2. \label{e6.10} 
\er
Solving these equations for $\G_{21}$ and $\G_{12}$, by considering the respective expansions in both positive and negative powers of $\lm$, we get
\br
 \G_{21}^{(0)} &=& \rmi e^{-\frac{\rmi\vp}{2}}, \quad \,\, \G_{12}^{(0)} \,=\, \rmi e^{\frac{\rmi\vp}{2}},  \quad\,\,
  {\hat\G}_{21}^{(0)} \,=\,\rmi e^{\frac{\rmi\vp}{2}}, \qquad {\hat\G}_{12}^{(0)}\,=\, \rmi e^{-\frac{\rmi\vp}{2}}, \\ 
 \G_{21}^{(1)}&=&-\frac{\rmi}{m}\left[\pa_t\vp + \pa_x\vp \right] e^{-\frac{\rmi\vp}{2}}, \qquad  {\hat\G}_{21}^{(1)} \,=\, \frac{\rmi}{m}\left[\pa_t\vp -\pa_x\vp\] e^{\frac{\rmi\vp}{2}},\\
 \G_{12}^{(1)} &=& -\frac{\rmi}{m}\left[\pa_t\vp + \pa_x\vp \right] e^{\frac{\rmi\vp}{2}}, \qquad\,\,\,
 {\hat\G}_{12}^{(1)} \,=\, \frac{\rmi}{m}\left[\pa_t\vp - \pa_x\vp \right] e^{-\frac{\rmi\vp}{2}},\\ 
 \G_{21}^{(2)} &=& e^{-\frac{\rmi\vp}{2}}\left[-\frac{2}{m^2}\,\pa_x\(\pa_t\vp+\pa_x\vp\) +\frac{\rmi}{2m^2}\(\pa_t\vp+\pa_x\vp\)^2 +\sin\vp \right], \\
 {\hat\G}_{21}^{(2)} &=&  e^{\frac{\rmi\vp}{2}}\left[-\frac{2}{m^2}\,\pa_x\(\pa_t\vp -\pa_x\vp\) +\frac{\rmi}{2m^2}\(\pa_t\vp -\pa_x\vp\)^2 -\sin\vp \right],\\
 \G_{12}^{(2)} &=& e^{\frac{\rmi\vp}{2}}\left[\frac{2}{m^2}\,\pa_x\(\pa_t\vp +\pa_x\vp\) +\frac{\rmi}{2m^2}\(\pa_t\vp+\pa_x\vp\)^2 -\sin\vp \right],\\
 {\hat\G}_{12}^{(2)} &=& e^{-\frac{\rmi\vp}{2}}\left[\frac{2}{m^2}\,\pa_x\(\pa_t\vp -\pa_x\vp\) +\frac{\rmi}{2m^2}\(\pa_t\vp-\pa_x\vp\)^2 +\sin\vp \right].
\er
Therefore, from the two generating functions of the infinite conserved quantities determined by (\ref{e6.7}) and (\ref{e6.8}), we have that the first non-vanishing conserved quantities are given by
\br
 {\tt I}_1^{(1)} = \frac{1}{4m\rmi}\int_{-\infty}^{\infty}\rmd x\left[\frac{1}{2}\(\pa_t\vp +\pa_x\vp\)^2 -m^2\cos\vp\right], \\
 {\hat{\tt I}}_1^{(1)} = \frac{\rmi}{4m}\,\int_{-\infty}^{\infty}\rmd x\left[\frac{1}{2}\(\pa_t\vp -\pa_x\vp\)^2 -m^2\cos\vp\right],\\
 {\tt I}_2^{(1)} = \frac{\rmi}{4m}\,\int_{-\infty}^{\infty}\rmd x\left[\frac{1}{2}\(\pa_t\vp +\pa_x\vp\)^2 -m^2\cos\vp\right],\\
 {\hat {\tt I}}_2^{(1)} =\frac{1}{4m\rmi}\int_{-\infty}^{\infty}\rmd x\left[\frac{1}{2}\(\pa_t\vp -\pa_x\vp\)^2 -m^2\cos\vp\right].
\er
Therefore, the usual energy and momentum for the sine-Gordon model are recovered by combining these results as follows,
\br
\fl  E&=&\rmi m \({\tt I}_1^{(1)}-{\tt I}_2^{(1)} - {\hat I}_{1}^{(1)} + {\hat I}_{2}^{(1)}\)\,=\, \int_{-\infty}^{\infty}\rmd x\left[\frac{1}{2}\bigg\{(\pa_t\vp)^2 +(\pa_x\vp)^2\bigg\} -m^2 \cos\vp\right],\\
\fl  P&=& \rmi m \({\tt I}_1^{(1)}-{\tt I}_2^{(1)} + {\hat I}_{1}^{(1)} - {\hat I}_{2}^{(1)}\)\,=\, \int_{-\infty}^{\infty}\rmd x\, (\pa_t\vp)(\pa_x\vp).
\er 

Now,  considering the generating function of infinite charges in the presence of a defect, we have that the first contributions to the modified conserved quantities are given by
\br
D_ 1 &=& - \ln \bigg[K_{11} + K_{12}\, {\G}_{21}\bigg]\bigg|_{x=0}, \label{e6.36.}
\er
where the defect matrix can be explicitly written as \cite{Bow2},
\br
 K = \left[
   \begin{array}{cc} 
    e^{-\frac{\rmi}{4}\(\vt-\vp \) }  & \lm^{-1} \,\s\,e^{-\frac{\rmi}{4}\(\vt+\vp \) } \\[0.3cm]
    -\lm^{-1}\,\s\,e^{\frac{\rmi}{4}\(\vt+\vp \) }  &  e^{\frac{\rmi}{4}\(\vt-\vp \) }
   \end{array}
 \].\label{e6.33.}
\er
Hence, taking into account the expansion in both negative and positive powers of $\lm$ and the form of the defect matrix (\ref{e6.33.}), we found that
\br
\fl {D}_{1}^{(1)} &=& - \rmi\s e^{-\rmi(\vt +\vp)/2}, \qquad {\hat D}_1^{(1)} \,\,=\,\, \frac{\rmi}{\s} e^{-\rmi(\vt-\vp)/2} - \frac{1}{m}\(\pa_t\vp-\pa_x\vp\).
\er
Now, following the same procedure for the second generating function, one gets
\br
 {D_2} &=& -\ln \bigg[K_{21}\G_{12} + K_{22}\bigg]\bigg|_{x=0},\label{e6.39.}
\er
from which we obtain the following contributions,
\br
\fl D_2^{(1)} &=&\rmi\s\, e^{\rmi(\vt +\vp)/2}, \qquad {\hat D}_2^{(1)}\,\,=\,\,-\frac{\rmi}{\s} \,e^{\rmi(\vt-\vp)/2} - \frac{1}{m}\(\pa_t\vp-\pa_x\vp\).
\er
Then, the corresponding defect energy and momentum are given by, 
\br
\fl E_D  &=& \rmi m\left[D_1^{(1)}  - D_2^{(1)} -{\hat D}_1^{(1)} +{\hat D}_2^{(1)} \right] = 2m \left[ \s \cos\(\frac{\vt +\vp}{2}\) + \frac{1}{\s }\cos\(\frac{\vt -\vp}{2}\)\right],\\[0.1cm]
\fl P_D  &=& \rmi m\left[D_1^{(1)}  - D_2^{(1)}  + {\hat D}_1^{(1)}  - {\hat D}_2^{(1)} \right] = 2m \left[ \s \cos\(\frac{\vt +\vp}{2}\) - \frac{1}{\s }\cos\(\frac{\vt -\vp}{2}\)\right],
\er
which are in complete agreement with the results previously obtained from both the Lagrangian \cite{Bow2} and the $r$-matrix \cite{Hab2} approach. 



\end{document}